\documentclass{INTERSPEECH2023}
\interspeechcameraready 
\usepackage{subfig,tabularx, url}

\title{DualVC: Dual-mode Voice Conversion using Intra-model Knowledge Distillation and Hybrid Predictive Coding\vspace{-5pt}}

\name{Ziqian Ning$^{1,2}$, Yuepeng Jiang$^{1}$, Pengcheng Zhu$^{2}$, Jixun Yao$^{1}$, Shuai Wang$^{3}$, Lei Xie$^{1*}$, Mengxiao Bi$^{2}$\thanks{* Corresponding author.}\vspace{-7pt}}

\address{
  $^1$Audio, Speech and Language Processing Group (ASLP@NPU), School of Computer Science, \\ Northwestern Polytechnical University, Xi'an, China\\
  $^2$Fuxi AI Lab, NetEase Inc., Hangzhou, China\\
  $^3$Shanghai Jiao Tong University, Shanghai, China\vspace{-2pt}}
\email{\{ningziqian, Jiangyp, yaojx\}@mail.nwpu.edu.cn, \{zhupengcheng, bimengxiao\}@corp.netease.com, wsstriving@gmail.com,lxie@nwpu.edu.cn}

\begin{document}
\maketitle
\begin{abstract}
\vspace{-2pt}

Voice conversion is an increasingly popular technology, and the growing number of real-time applications requires models with streaming conversion capabilities. Unlike typical (non-streaming) voice conversion, which can leverage the entire utterance as full context, streaming voice conversion faces significant challenges due to the missing future information, resulting in degraded intelligibility, speaker similarity, and sound quality. To address this challenge, we propose \textit{DualVC}, a dual-mode neural voice conversion approach that supports both streaming and non-streaming modes using jointly trained separate network parameters. Furthermore, we propose intra-model knowledge distillation and hybrid predictive coding (HPC) to enhance the performance of streaming conversion.
Additionally, we incorporate data augmentation to train a noise-robust autoregressive decoder, improving the model's performance on long-form speech conversion. Experimental results demonstrate that the proposed model outperforms the baseline models in the context of streaming voice conversion, while maintaining comparable performance to the non-streaming topline system that leverages the complete context, albeit with a latency of only 252.8 ms.
\end{abstract}
\vspace{-2pt}
\section{Introduction}


\begin{figure*}[!htbp]
\centering
\includegraphics[scale=0.45]{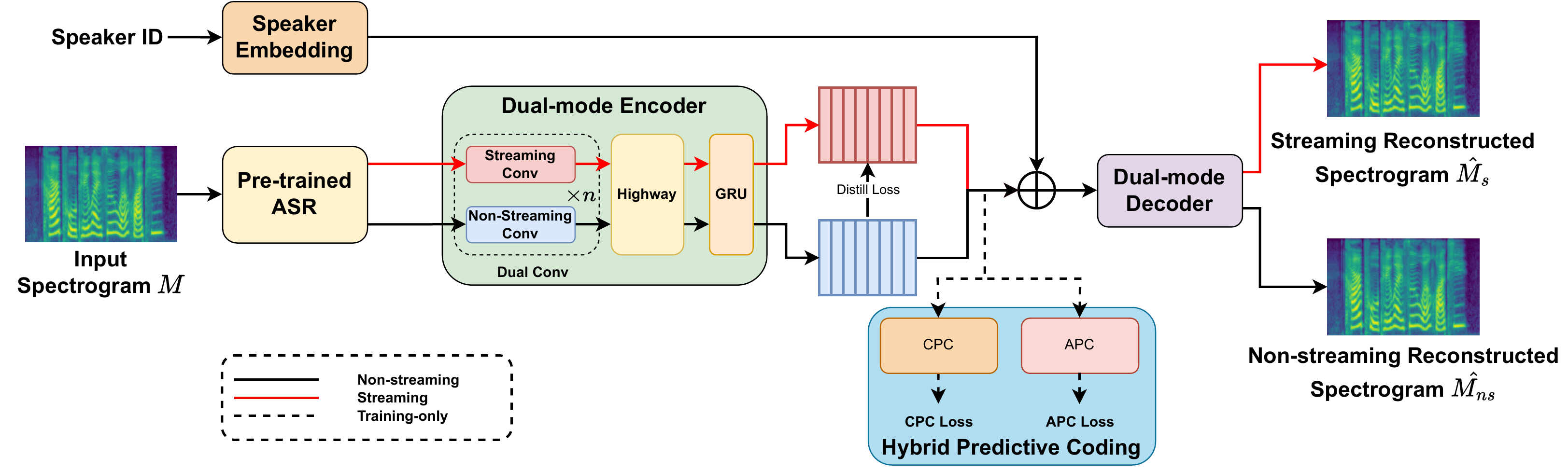}
\caption{The architecture of DualVC}
\label{fig:model}
\end{figure*}
Voice conversion (VC) is a technique that transforms a speaker's voice into that of another speaker without altering the linguistic content~\cite{DBLP:journals/taslp/SismanYKL21}. The advances of deep learning have significantly contributed to the rapid development of voice conversion, evolving the capabilities of generating natural-sounding speech. VC has been extensively applied in diverse applications including privacy protection~\cite{DBLP:journals/corr/abs-2211-03038} and movie dubbing~\cite{DBLP:journals/corr/abs-2201-00269}. However, the increasing diversity of VC applications, such as live broadcasting and other real-time communication (RTC) applications, has led to higher demands for streaming capabilities.

While non-streaming VC models~\cite{DBLP:conf/interspeech/WangDYCLM21,DBLP:conf/interspeech/WangZYLDXGCL21,DBLP:conf/interspeech/LiZM21,DBLP:conf/icml/QianZCYH19} have demonstrated impressive conversion quality, they require full-utterance input and are not feasible for real-time applications. 
In contrast, despite the sustained efforts in developing streaming VC models, their performance still falls short when compared to non-streaming models. This is mainly due to the challenges associated with processing chunked or framewise input on the fly and without access to future information.
Consequently, its performance may suffer from low intelligibility, poor sound quality, and inferior speaker similarity. 

One promising approach that may alleviate these problems is to use Intermediate Bottleneck Features (IBF), as discussed in~\cite{DBLP:journals/corr/abs-2210-15158}. Instead of using bottleneck features (BNF)~\cite{DBLP:conf/icmcs/SunLWKM16,DBLP:conf/icassp/ZhaoLSWKTM22} from the output of a pre-trained Automatic Speech Recognition (ASR) encoder, IBF is extracted from the middle layers of the ASR encoder, with the premise to preserve more information to compensate for mispronunciations caused by streaming ASR with degraded performance. However, IBF contains more timbre of the source speaker, resulting in timbre leakage. Apart from enhancing input features, there are alternative approaches that aim to improve the streaming voice conversion performance from a model structure perspective.
Yang et al.~\cite{DBLP:conf/interspeech/YangDYZX22} disentangle speaker timbre and linguistic content by leveraging vector quantization (VQ)~\cite{DBLP:conf/interspeech/WuCL20}, mutual information (MI)~\cite{DBLP:conf/icml/ChengHDLGC20} minimization, and contrastive predictive coding (CPC)~\cite{DBLP:journals/corr/abs-1807-03748}, and leverages BNF as additional input to enhance intelligibility. FastS2S-VC~\cite{DBLP:journals/corr/abs-2104-06900} developed a non-autoregressive sequence-to-sequence model with a novel attention predictor.

To narrow the performance gap between non-streaming and streaming VC systems, a prevalent strategy is to leverage the non-streaming system as a teacher to provide guiding signals to augment the streaming counterpart~\cite{DBLP:journals/corr/abs-2210-15158,  DBLP:journals/corr/abs-2104-06900, DBLP:conf/icassp/HayashiKT22}. Despite the effectiveness exhibited by this teacher guidance, the majority of existing approaches rely on a separate pre-trained voice conversion model, thus resulting in a more complicated pipeline.

In this paper, we propose \textit{DualVC}, a novel dual-mode VC model that supports both streaming and non-streaming inference. Instead of introducing a pre-trained non-streaming teacher model, we employ dual-mode convolution to unify the non-streaming teacher and the streaming student into a single model  and perform joint optimization. Despite the similar dual-mode joint training approach, it is worth noting that our \textit{DualVC} model differs from~\cite{DBLP:conf/icassp/HayashiKT22} in the following respects.
1) Unlike the approach in~\cite{DBLP:conf/icassp/HayashiKT22}, where the parameters for non-streaming and streaming modes are shared and have mutual effects, our model employs distinct parameters for different modes.
2) While the distillation loss is directly computed between the model outputs in~\cite{DBLP:conf/icassp/HayashiKT22}, our model calculates the loss between intermediate features. 
3) Additionally, we detach the non-streaming module during the distillation process to avoid potential interference from the student model on the teacher model.

Furthermore, we introduce a hybrid predictive coding (HPC) mechanism to compensate for the absence of future information in the streaming mode. HPC integrates contrastive and autoregressive predictive coding methods~\cite{DBLP:journals/corr/abs-1807-03748,DBLP:conf/interspeech/ChungHTG19}, and encourages the encoder to learn a more resilient feature structure in cases where future information is unavailable.

Finally, to alleviate the problem of error accumulation in long-sentence inference without future information, we introduce noise to both the input Mel-spectrogram and the gradient of the autoregressive module. Extensive experiments demonstrate that the proposed streaming DualVC outperforms the baseline system, achieving similar conversion quality to the non-streaming topline system with a latency of only 252.8 ms.
\section{Proposed Approach}

As illustrated in Fig.~\ref{fig:model}, DualVC is built on a recognition-synthesis framework, comprising an encoder, a HPC module, and a decoder. Initially, the encoder of a pre-trained ASR model extracts BNF from the input spectrogram. These BNFs are then forwarded to the encoder to further extract contextual information. The HPC module, which is only used during the training phase, facilitates the encoder in extracting more effective latent representation via unsupervised learning methods. Subsequently, the target speaker embedding is concatenated to the latent representation and provided as input to the decoder. Finally, the decoder generates the converted spectrogram. 

\vspace{-2pt}
\subsection{Streamable Architecture}
\vspace{-2pt}

The backbone of DualVC is CBHG-AR \cite{tian2020nus}, which consists of a CBHG ~\cite{DBLP:conf/icml/Skerry-RyanBXWS18} encoder and an autoregressive (AR) decoder. To enable streaming, any components that rely on future information must be modified or replaced. Specifically, bidirectional GRU layers and convolution layers are replaced with unidirectional GRU layers and causal convolution layers, respectively. In typical convolutional neural networks, paddings are added to both sides of the input to ensure equal lengths of input and output features. Thus, the convolutional kernel is able to access historical, current, and future information within its perceptual field. However, during the streaming inference, future information is not available, demanding the usage of causal convolution layers with all padding shifted to the left of the input, which involves no future information. With unidirectional GRU layers, the model can infer without relying on any future information, accepting only the current frame and the last hidden states as input.

\vspace{-2pt}
\subsection{Dual-mode Convolution}
\vspace{-2pt}

The utilization of causal convolution introduces the disadvantage of missing future information, leading to degraded performance. To address this issue,
we propose to use dual-mode convolution in couples with intra-model knowledge distillation.
 
 

In our proposed model, we adopt a modified variant of the depthwise separable convolution ~\cite{DBLP:conf/cvpr/Chollet17} as the basic convolution layer, in which a depthwise convolution layer is sandwiched between two pointwise convolution layers, followed by a dropout layer at the end.
The dual-mode convolution block consists of two parallel basic convolution layers, one of which is causal for streaming mode and the other non-causal for non-streaming mode. 
All convolution layers in the backbone model are replaced with the dual-mode convolution block, and we forward the model twice using two modes respectively during training. To bring the streaming intermediate representation closer to the non-streaming one, we calculate the loss between the streaming encoder output $Z$ and the non-streaming encoder output $\hat{Z}$. Since both modes are trained together without relying on a pre-trained teacher model, we refer to this process as intra-model knowledge distillation. The knowledge distillation loss is formulated as
\begin{equation}
\mathcal{L}_{distill} = \text{SmoothL1Loss}(Z, detach(\hat{Z})),
\end{equation}
\vspace{-2pt}
where the SmoothL1Loss is the smoothed version of L1 loss defined in Fast R-CNN~\cite{girshick2015fast} and measures the element-wise difference between $Z$ and $\hat{Z}$.
$\hat{Z}$ is detached in order to bring the output of streaming mode closer to non-streaming mode, without affecting non-streaming mode.

Dual-mode convolution not only allows a single model to be used for both streaming and non-streaming scenarios, but also enhances the performance of the streaming convolution by using the output of non-streaming convolution as guidance during training. 
Dual-mode models have been investigated in the literature. Dual-model ASR models~\cite{DBLP:conf/iclr/YuHGCLSWP21,zhang2020unified} employ dynamic chunk size to train monotonic transformer modules and use shared weight to perform causal and non-causal convolutions, improving the performance in both streaming and non-streaming cases.
A previous study on streaming voice conversion~\cite{DBLP:conf/icassp/HayashiKT22} also adopts the weight-sharing strategy, while the convolutional kernel parameters with future receptive fields are discarded during streaming inference by shrinking kernel size.
In contrast, our approach utilizes distinct convolutional layer parameters for streaming and non-streaming respectively. 
As either streaming or non-streaming convolution is selectively utilized during inference, the extra parameters incur zero computation overhead. Regarding the dual-mode decoder, distillation loss is unnecessary since its output is directly bounded by the ground-truth Mel-spectrogram.

\begin{figure}[ht]
 
\centering
\includegraphics[scale=0.8]{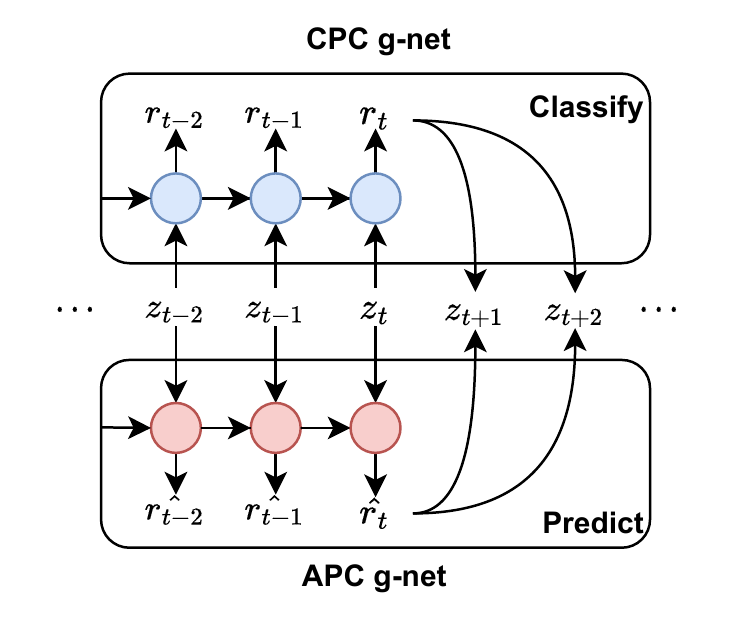}
\caption{Hybrid Predictive Coding Module}
\label{fig:hpc}
\vspace{-10pt}
\end{figure}
\vspace{-2pt}

\subsection{HPC for Unsupervised Latent Representation Learning}

Besides using dual-mode convolution with intra-model knowledge distillation, we aim to further improve the conversion quality by enhancing the latent representation extracted by the CBHG encoder. 
To this end, we propose HPC consisting of CPC~\cite{DBLP:journals/corr/abs-1807-03748} and Autoregressive Predictive Coding (APC)~\cite{DBLP:conf/interspeech/ChungHTG19}, which are unsupervised representation learning methods.

CPC employs an autoregressive-based g-net to extract the aggregation $R_k = \{r_{k, 1}, r_{k, 2}, \cdots\}$ from $Z$ and is trained with InfoNCE loss~\cite{DBLP:conf/apsipa/OyamadaKKAHK17}. By using $r_{k,t}$ as input, the g-net distinguishes positive from negative samples in future m steps $[z_{k,t+1}, z_{k,t+m}]$, thus encouraging the latent representation $Z$ to capture better feature structure. 
Different negative sample selection methods affect what is encoded in the representation, thus the important parts of the features can be extracted and unwanted parts can be discarded using prior knowledge.
On the other hand, APC is an autoregressive model that predicts $[z_{k,t+1}, z_{k, t+m}]$ directly by minimizing the L1 loss. Contrary to CPC, APC does not use prior knowledge to select negative samples for representation learning, allowing it to preserve more information with better flexibility.

To leverage the strengths of both CPC and APC, we propose a hybrid approach called HPC, which is shown in Fig.~\ref{fig:hpc}. HPC adopts separate g-nets for CPC and APC, and performs classification between positive and negative samples as well as straight predictions. 
Although future information cannot be acquired during streaming inference, the common feature structure captured by the HPC module allows the model to infer its content to some extent.

The HPC loss can be formulated as:

\begin{equation}
\mathcal{L}_{HPC} = L_{CPC} + L_{APC}
\end{equation}
In the experiment section, we will show that HPC provides a more comprehensive and robust representation of the content information, improving the effectiveness of the streaming voice conversion.
\subsection{Noise Robust Autoregressive Decoder}
We utilize an autoregressive structure for the decoder due to its exceptional generative capabilities, and it can generate based solely on historical information as input. 
However, we observe that the conversion quality tends to decline over time due to the accumulation of errors in the autoregressive process for upstream features that have already incurred losses. This challenge is especially severe in streaming models where the input audio is considered to have infinite length. Hence, it is crucial to enhance the robustness of the AR structure. 

In this paper, we propose a novel data augmentation approach that combines input feature augmentation and gradient augmentation to reduce the mismatch between the low-quality Mel-spectrogram of the actual input and the high-quality ground truth Mel-spectrogram used in training. By adding normally distributed noise $n \in \mathcal{N}(0, 1)$ to the ground truth input of the AR, we reintroduce features with errors in the autoregressive process in the following step. Also, noise $\hat{n} \in \mathcal{N}(0, 10^{-6})$ is added to the gradient of the AR module during training to further improve robustness.

The overall objective function consists of three parts: $L_{distill}$ and $L_{HPC}$ are described above, and $L_{rec}$ is the reconstruction loss calculated between ground truth Mel-spectrogram $Y$ and generated one $\hat{Y}$\footnote{Note that we compute $L_{HPC}$ and $L_{rec}$ for both the streaming and non-streaming mode}.
\begin{equation}
\mathcal{L}_{rec} = \text{MSELoss}(Y, \hat{Y})
\end{equation}
\begin{equation}
\mathcal{L} = L_{distill} + L_{HPC} + L_{rec}
\end{equation}

\begin{table*}[!htbp]
\caption{Comparison of the dual-mode DualVC with IBF-VC, topline, bottomline, and ablation models regarding speaker similarity and speaker naturalness MOS with confidence intervals of 95\% under 2 voice conversion scenarios. NMOS denotes naturalness MOS, and SMOS denotes speaker similarity MOS. A higher value means better performance.}
\vspace{-4pt}
 \label{tab:mos}
\setlength{\tabcolsep}{3mm}
 \centering
 \resizebox{\linewidth}{!}{
 \renewcommand{\arraystretch}{1.2}
\begin{tabular}{l|ccc|ccc|ccc}
\hline
           & \multicolumn{3}{c|}{Clean}                           & \multicolumn{3}{c|}{Noisy}                           & \multicolumn{3}{c}{Overall}                          \\ \hline
           & NMOS $\uparrow$ & SMOS $\uparrow$ & CER(\%) $\downarrow$ & NMOS $\uparrow$ & SMOS $\uparrow$ & CER(\%) $\downarrow$ & NMOS $\uparrow$ & SMOS $\uparrow$ & CER(\%) $\downarrow$ \\ \hline
Topline    & 3.98$\pm$0.03   & 3.89$\pm$0.04   & 8.7              & 3.83$\pm$0.02   & 3.80$\pm$0.05   & 10.1             & 3.91$\pm$0.02   & 3.84$\pm$0.02   & 9.4              \\
IBF-VC~\cite{DBLP:journals/corr/abs-2210-15158}     & 3.79$\pm$0.03   & 3.76$\pm$0.02   & 12.7             & 3.68$\pm$0.03   & 3.71$\pm$0.02   & 14.9             & 3.78$\pm$0.05   & 3.74$\pm$0.05   & 13.8             \\
Bottomline & 3.32$\pm$0.04   & 3.57$\pm$0.02   & 20.9             & 3.21$\pm$0.03   & 3.52$\pm$0.04   & 23.2             & 3.26$\pm$0.03   & 3.55$\pm$0.03   & 22.0             \\ \hline
DualVC (non-streaming) & 4.04$\pm$0.03   & 3.90$\pm$0.03   & 8.2              & 3.87$\pm$0.04   & 3.82$\pm$0.02   & 9.8             & 3.96$\pm$0.03   & 3.86$\pm$0.02   & 9.0\\
DualVC (streaming)     & 3.83$\pm$0.03   & 3.81$\pm$0.03   & 10.3             & 3.76$\pm$0.04   & 3.74$\pm$0.03   & 11.4             & 3.80$\pm$0.03   & 3.81$\pm$0.02   & 10.9             \\
\hspace{1em}-Dual mode & 3.44$\pm$0.04   & 3.70$\pm$0.04   & 17.2             & 3.37$\pm$0.03   & 3.60$\pm$0.04   & 18.9             & 3.41$\pm$0.05   & 3.65$\pm$0.04   & 18.0             \\
\hspace{1em}-CPC       & 3.74$\pm$0.05   & 3.78$\pm$0.02   & 11.6             & 3.65$\pm$0.04   & 3.70$\pm$0.03   & 13.0             & 3.70$\pm$0.04   & 3.74$\pm$0.02   & 12.3             \\
\hspace{1em}-APC       & 3.71$\pm$0.03   & 3.76$\pm$0.04   & 12.0             & 3.63$\pm$0.03   & 3.73$\pm$0.04   & 13.7             & 3.67$\pm$0.02   & 3.75$\pm$0.03   & 12.9             \\
\hspace{1em}-HPC       & 3.68$\pm$0.04   & 3.75$\pm$0.04   & 12.8             & 3.53$\pm$0.03   & 3.64$\pm$0.04   & 14.5             & 3.61$\pm$0.02   & 3.70$\pm$0.03   & 13.7             \\
\hspace{1em}-AR noise  & 3.65$\pm$0.04   & 3.76$\pm$0.04   & 13.4             & 3.52$\pm$0.02   & 3.67$\pm$0.05   & 16.1             & 3.59$\pm$0.05   & 3.72$\pm$0.05   & 14.8             \\ \hline
\end{tabular}
}
\vspace{-15pt}
\end{table*}
\vspace{-5pt}
\section{Experiments}
\vspace{-1pt}
In the experiments, all testing VC models were trained on an internal Mandarin corpus, containing $20000$ neutral utterances uttered by $20$ speakers, with each speaker contributing $1000$ utterances. One male and one female speaker were reserved as the target speakers for voice conversion tests. $10$ clean and $10$ noisy clips are used as source recordings. The selected recordings were then converted to the two target speakers using the proposed model and all comparison models to further perform evaluations. All the speech utterances are resampled to 16 kHz. Besides, tempo augmentation was adopted to enrich prosody diversity~\cite{DBLP:conf/icassp/ZhaoLSWKTM22}, using a random multiplier of 0.8-1.5. During training, augmented and original features were fed to the VC model alternatively~\cite{DBLP:journals/corr/abs-2211-04710}.

Mel-spectrogram and BNF were computed at a frame length of 50ms and a hop size of 12.5ms. The ASR system for BNF extraction was Fast-U2++ ~\cite{DBLP:journals/corr/abs-2211-00941} implemented by WeNet toolkit~\cite{DBLP:conf/interspeech/YaoWWZYYPCXL21}, and trained on a Mandarin ASR corpus Wenetspeech~\cite{DBLP:conf/icassp/ZhangLGSYXXBCZW22}. To reconstruct waveform from the converted Mel-spectrograms, we use DSPGAN ~\cite{DBLP:journals/corr/abs-2211-01087}, which is a robust universal vocoder based on the time-frequency domain supervision from digital signal processing (DSP).

To evaluate the performance of the proposed model in streaming voice conversion, \textbf{IBF-VC} ~\cite{DBLP:journals/corr/abs-2210-15158}, which is also built on the recognition-synthesis framework, was selected as the baseline system. Since IBF-VC is improved in terms of input features and knowledge distillation, there are no specific requirements for the structure of the model itself, we also use CBHG-AR as the backbone to reimplement IBF-VC for a fair comparison. The base CBHG-AR model with no modification and use full-utterance input is treated as the topline model, while we simply replace the convolution layers with the causal version to form a naive streaming implementation as the bottomline model. As a dual-mode model, both streaming and non-streaming modes of DualVC were evaluated in the experiments.

\subsection{Subjective Evaluation}
We conducted Mean Opinion Score (MOS) tests to evaluate the naturalness and speaker similarity of different models. The naturalness metric mainly considers intelligibility, prosody, and sound quality. A higher naturalness MOS score indicates the converted speech sounds more human-like. In both MOS tests, there are 20 listeners participated. Particularly for the speaker similarity test, we use the target speaker's real recording as the reference. We recommend the readers listen to our samples\footnote{Demo: https://dualvc.github.io/}.

\subsubsection{Speech Naturalness}
The NMOS results presented in Table~\ref{tab:mos} indicate that our proposed DualVC can achieve the best performance in speech naturalness. Specifically, with clean input, streaming DualVC outperforms IBF-VC and achieves a MOS score close to the topline model. The non-streaming DualVC even exceeds the topline model with the additional HPC module, showing the strong capability of unsupervised learning strategies. While all models show different degrees of performance degradation when accepting the noisy input, DualVC demonstrates minimal degradation, proving its superior robustness.

\subsubsection{Speaker Similarity}
The results of SMOS tests across different models are also shown in Table 1. In accordance with the naturalness metrics, for the speaker similarity, the streaming DualVC technique attained results that were inferior only to the top-performing model. Although a decrease in performance was noticed for the noise input condition, DualVC sustained its standing.
In light of the speaker similarity and naturalness performance, the DualVC method exhibits a remarkable superiority for streaming voice conversion.

\subsubsection{Ablation Study}
To investigate the importance of our proposed methods in DualVC, three ablation systems were obtained by dropping dual-mode convolution, HPC module and noise-augmented training of autoregressive decoder. These systems are referred to as \textit{-Dual-mode Conv}, \textit{-HPC} and \textit{-AR Noise}, respectively. Note that the knowledge distillation loss is also discarded when we get rid of the dual-mode convolution. 
To demonstrate the advantages of combining CPC and APC, we carried out additional experiments where either CPC or APC is omitted, denoted as \textit{-CPC} and \textit{-APC}, respectively.
As shown in Table 1, the removal of these methods brings obvious performance decreases with respect to both speech naturalness and speaker similarity. 
Notably, the elimination of the dual-mode convolution brings the most significant decline, approaching the performance of the bottom line. This observation demonstrates that the non-streaming model is an extremely strong guide for the streaming model.

\subsection{Objective Evaluation}

\subsubsection{Intelligibility Evaluation}
We utilized the same pre-trained ASR model for extracting BNF and recognizing the source speech, converted clean, and noisy clips. To obtain more accurate results, a larger set of 200 samples was tested. The character error rate (CER) is also reported in Table~\ref{tab:mos}. The CER for source speech is 6.0\% and 8.4\% for clean and noisy clips, respectively. We can see that the bottomline obtains the highest CER, indicating bad intelligibility. In contrast, streaming DualVC achieves a CER close to the topline, and both systems induce a small CER increase compared to the source speech, demonstrating its ability to maintain good intelligibility despite the lack of future information.

\begin{table}[]
\centering
 \caption{Computation \& Real-time Metrics of DualVC.}
 \vspace{-4pt}
\setlength{\tabcolsep}{0.6mm}
 \label{tab:performance}
\setlength{\tabcolsep}{3mm}{
\begin{tabular}{l|ccc}
\hline
        & RTF                      & Latency (ms)               & FLOPs (G)                \\ \hline
ASR     & 0.26                     & 41.6                      & 8.4                 \\
DualVC  & 0.12                     & 19.2                      & 4.7              \\
Vocoder & 0.20                     & 32.0                      & 5.2                 \\ \hline
All     & 0.58                     & 92.8                      & 18.3                   \\ \hline
\end{tabular}
}
\vspace{-10pt}
\end{table}

\subsubsection{Computational Efficiency Evaluation}
In this study, we considered three major metrics to assess performance: real-time factor (RTF), latency, and floating point ops (FLOPs). The results are shown in Table ~\ref{tab:performance}.
RTF is a common measure of model inference speed that expresses the ratio between model inference time and input feature duration. To meet real-time requirements, the RTF needs to be less than 1, and the RTF of our complete pipeline on a single Intel Xeon Silver 4210 core was 0.58. Latency is defined as the interval between the time of user input and model output, which consists of three parts: model inference, input waiting, and network latency. With network latency not taken into account, the system latency can be expressed as
\begin{equation}
Latency = chunksize \times (1 + RTF).
\end{equation}
With a chunk size of 160 ms and a model inference latency of 92.8 ms, the total latency was calculated to be 252.8 ms. FLOPs were used to quantify the computational complexity of the model, and our DualVC model had a FLOP value of 5.2 G, whereas the one for the whole pipeline was 18.3 G.

\section{Conclusions}
In this paper, we proposed a dual-mode voice conversion model (DualVC) to address the challenge of limited future information in real-time applications. The DualVC model utilizes dual-mode convolution with intra-model distillation, and hybrid predictive coding consisting of CPC and APC for unsupervised representation learning to enhance the conversion quality. Experiments showed that streaming DualVC outperformed the baseline system and achieved similar performance to the topline system with a latency of only 252.8 ms.

\bibliographystyle{IEEEtran}

\bibliography{mybib}

\end{document}